\documentclass{PoS}
\usepackage{graphicx}

\title{Recent results on collective effects in small systems from PHENIX at RHIC}

\ShortTitle{Collective effects in small systems from PHENIX at RHIC}

\author{\speaker{Sarah Campbell for the PHENIX Collaboration}\\
        Columbia University\\
        E-mail: \email{sc3877@columbia.edu}}

\abstract{Collisions of simple systems, such as $p$+$p$, or $p$+Nucleus have been used as benchmarks for our understanding of heavy ion collsions, since it was assumed they would be free of the effects from hot nuclear matter. \\ \\
Recently long range correlations and anisotropies of momentum spectra have been seen in such collisions, challenging this assumption.  Such phenomena have been understood to be the result of the collective motion, which can best be described by hydrodynamics, whose initial conditions are set by the geometry of the colliding systems, together with their fluctuations.  This talk will discuss the recent results from the PHENIX experiment at RHIC using a variety of colliding species ($p$+Au, $d$+Au, $^{3}$He+Au) that give a better understanding of the origin of the observed correlations and anisotropies, thus providing insight as to whether a quark gluon plasma is formed in these simple systems.}

\FullConference{38th International Conference on High Energy Physics\\
		3-10 August 2016\\
		Chicago, USA}

\begin{document}

\section{Introduction}

Long range hadron correlations, also known as ridge phenomena, were first measured in heavy ion collisions by the PHOBOS~\cite{PHOBOS} and STAR~\cite{STAR} collaborations at the Relativistic Heavy Ion Collider (RHIC) at Brookhaven National Laboratory. Dihadron correlation functions in $\Delta \phi$ with a large separation in rapidity are decomposed into cosine Fourier component amplitudes, $c_{n}$.  These pair anisotropy factors, $c_{n}$, are then separated into independent single particle azimuthal anisotropies, $v_{n}$, according to $c_{n}(p_{T1}, p_{T2}) = v_{n}(p_{T1}) \times v_{n}(p_{T2})$.~\cite{Luzum}  In heavy ion collisions, $v_{2}$ at low $p_{T}$ are seen as the result of collective behavior, often called flow, specific to high-temperature quark-gluon plasma (QGP) created in the collision.  Similarly-sized $v_{n}$ are measured in $\sqrt{s_{NN}}=200$~GeV Au+Au collisions at RHIC and in $\sqrt{s_{NN}}=2.76$~TeV Pb+Pb collisions at the Large Hadron Collider (LHC) at CERN.  The RHIC and LHC anisotropies are well described by the same hydrodynamic model for $v_{n}$ values up to the fifth-order.~\cite{GaleHydro}  
While it is unclear whether hydrodynamic flow of the QGP or alternative theories, such as initial-state gluon saturation,~\cite{GLASMA} cause these anistropies, the similarity in RHIC and LHC $v_{n}$ suggests that the same underlying processes drive the collectivity in these heavy ion collisions despite the large difference in collision energy.

At RHIC and the LHC, $p$+A or $d$+A collisions serve as a reference to control for effects stemming from the initial binding of quarks and gluons inside the nucleus as opposed to QGP behavior generated in A+A collisions.  However, in 2013, CMS discovered a long-range dihadron correlations in $\sqrt{s_{NN}}=5.02$~TeV $p$+Pb collisions at the LHC.~\cite{CMS}  This was confirmed in subsequent LHC measurements by the ATLAS~\cite{ATLAS} and ALICE~\cite{ALICE} experiments.  These results raise the question whether small system anisotropies are produced by the same processes as in heavy ion collisions and even if a QGP is being generated in the collision of small systems.

This proceeding presents azimuthal anisotropies in $\sqrt{s_{NN}}=200$~GeV $p$+Au, $d$+Au and $^{3}$He+Au collisions measured by the PHENIX experiment at RHIC.  This variation among small systems is a unique capability of the RHIC facility.  These measurements not only confirm that substantial azimuthal anisotropies are present in small systems at RHIC, but also allow us to test whether the $v_{n}$ in small systems vary with the initial collision geometry, a property of hydrodynamics.  First, the $d$+Au measurements are presented, followed by $p$+Au and $^{3}$He+Au results.

\section{$v_{n}$ measurements in $d$+Au}

PHENIX was the first RHIC experiment to observe azimuthal anisotropies from rapidity separated dihadron correlations in $\sqrt{s_{NN}}=200$~GeV $d$+Au collisions.  This was done using hadrons at mid-rapidity with a rapidity separation, $|\Delta \eta|$, between $0.45$ and $0.7$.~\cite{ppg149}  Here, PHENIX has expanded on this work with the measurement of long-range correlations in $\sqrt{s_{NN}}=200$~GeV $d$+Au using mid-rapidity hadrons and clusters in the Au-going Muon Piston Calorimeter (MPC).  This has the benefit of a larger $|\Delta \eta|$ between $2.75$ and $4.05$.~\cite{dAudeta}  By also considering the correlation of mid-rapidity $\pi^{0}$'s with clusters in the Au-going MPC, we extend these measurement out to higher $p_{T}$ values of $8$~GeV/$c$.~\cite{TakaoQM}  In both the hadron-MPC cluster and $\pi^{0}$-MPC cluster measurements, the azimuthal correlations exhibits a centrality dependence with larger anisotropies in central $d$+Au collisions.  The higher $p_{T}$ $\pi^{0}$-MPC cluster measurements are particularly promising because in heavy ion collisions azimuthal asymmetries at high $p_{T}$ are associated with the path length dependence of energy loss phenomena in the QGP.  While the $\pi^{0}$-MPC cluster correlations are still in their preliminary stage, they present a new way to study the possibility of energy loss in $d$+Au collisions.

\begin{figure}
\begin{center}
\includegraphics[width=.6\textwidth]{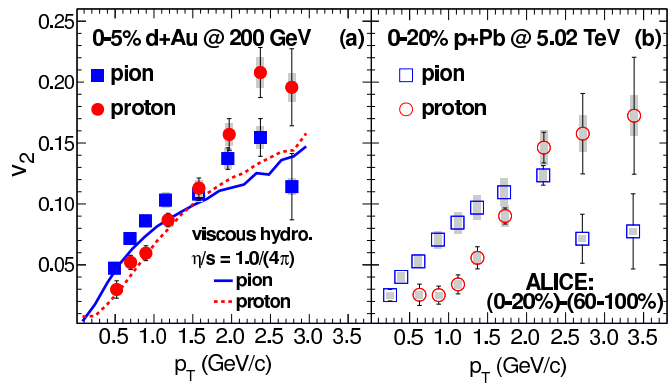}
\caption{Measured $v_{2}$ for identified pions and protons in (a) $0$-$20\%$ $d$+Au collisions at $200$~GeV at RHIC and (b) $0$-$20\%$ $p$+Pb collisions at $5.02$~TeV the LHC.~\cite{dAudeta}  The $d$+Au results are compared to a viscous hydrodynamic calculation.~\cite{LuzumRomatschke}~\cite{SONIC}}
\label{Fig:v2dAupid}
\end{center}
\end{figure}

PHENIX has also measured the proton and pion second-order azimuthal anisotropies, $v_{2}$, in $d$+Au using the event-plane technique.   With this method, the $v_{2}(p_{T})$ is defined as the average of the single particle's $cos 2(\phi - \Psi_{EP})$ value for a given $p_{T}$ bin divided by the event-plane resolution.  The event-plane, $\Psi_{EP}$, is determined from the Au-going MPC detector and the pions and protons are identified in the Time of Flight detector at mid-rapidity.~\cite{dAudeta}  Figure~\ref{Fig:v2dAupid} presents the pion and proton $v_{2}$ in (a) $0$-$5\%$ $d$+Au collisions at $\sqrt{s_{NN}}=200$~GeV measured by PHENIX and (b) $0$-$20\%$ $p$+Pb collisions at $\sqrt{s_{NN}}=5.02$~TeV measured by ALICE using pair correlations with the peripheral event subtraction method.~\cite{ALICE}  The RHIC and LHC results are of similar size and both show a clear mass ordering.  Mass ordering of $v_{2}$ at low $p_{T}$ is also seen in heavy ion collisions and is expected if the source is hydrodynamic.  Additionally, the $d$+Au pion and proton results are well described by a viscous hydrodynamic model.~\cite{LuzumRomatschke}~\cite{SONIC}

\subsection{$^{3}$He+Au and $p$+Au}

The analysis of the more recent $p$+Au and $^{3}$He+Au data benefits from upgrades to the PHENIX detector.  The forward vertex detector (FVTX) is a silicon tracking detector with $|\eta| \in [1,3]$ and a high multiplicity trigger uses the FVTX and allows almost all central events to be recorded, roughly a factor of ten increase over what would be collected by the minimum bias trigger.

\begin{figure}
\begin{center}
\includegraphics[width=.6\textwidth]{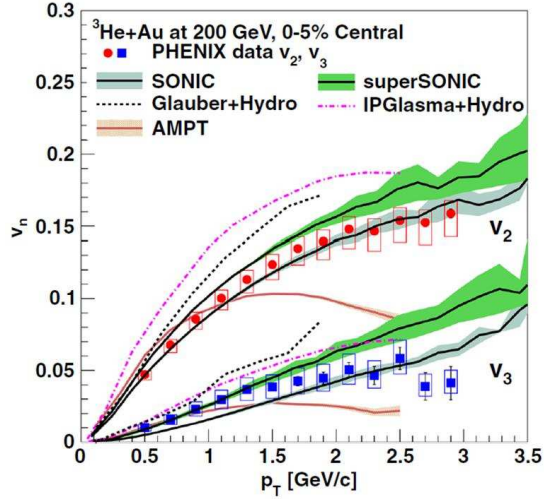}
\caption{The charged hadron $v_{2}$ and $v_{3}$ in $0$-$5\%$ $^{3}$He+Au collisions at $200$~GeV at RHIC.~\cite{He3Au}  The measurements are compared to theory calculations including SONIC~\cite{SONIC}, a Glauber model initial state with a hydrodynamic evolution~\cite{Glauber}, AMPT~\cite{AMPT}, SuperSONIC~\cite{superSONIC} and an IP-Glasma initial state with a hydrodynamic evolution.~\cite{IPGlasma}}
\label{Fig:vnHe3Au}
\end{center}
\end{figure}

The nuclear overlap in $^{3}$He+Au collisions has an intrinsic triangular initial state geometry coming from the shape of the $^{3}$He.  If the initial state geometry plays a role in the $v_{n}$, as it does in hydrodynamic models, positive $v_{3}$ values are expected. 
The charged hadron $v_{2}$ and $v_{3}$ in $\sqrt{s_{NN}}=200$~GeV $^{3}$He+Au are measured using the event-plane method.  
Figure~\ref{Fig:vnHe3Au} shows the charged hadron $v_{2}$ and $v_{3}$ in $0$-$5\%$ $^{3}$He+Au collisions where the event-plane is determined by the FVTX.  The measured $v_{3}$ values are positive for all $p_{T}$'s.  This is consistent with expectations from the initial triangular collision geometry for $^{3}$He+Au.~\cite{He3Au}  The $v_{2}$ and $v_{3}$ values are compared to various theories including SONIC~\cite{SONIC}, a Glauber model initial state with a hydrodynamic evolution~\cite{Glauber}, AMPT~\cite{AMPT}, superSONIC~\cite{superSONIC} and an IPGlasma initial state with a hydrodynamic evolution.~\cite{IPGlasma}  Four of these theories include a viscous hydrodynamic contribution.  The fifth model, AMPT, uses partonic and hadronic rescattering to generate the anisotropies but underpredicts the values particularly at high $p_{T}$.  The SONIC model, which uses a Glauber initial condition and a hadronic cascade, best describes the data, particularly at higher $p_{T}$. At $p_{T}<1.5$~GeV/$c$, the superSONIC prediction, which extends the SONIC model to include pre-equilibrium effects, also describes the data well.

The $v_{2}$ of identified particles ($\pi$, K, p) in $^{3}$He+Au collisions are also measured using the event plane method.  They show a clear mass-ordering at low $p_{T}$ and meson-baryon splitting at $p_{T}$ above $1.5$~GeV/$c$.  When the $v_{2}$ is scaled by the number of constituent quarks, $n_{q}$, a uniform curve is measured for each particle type as a function of the $n_{q}$-scaled transverse kinetic energy, $KE_{T}/n_{q}$. This mimics the phenomena of $n_{q}$-scaling seen in heavy ion collisions.~\cite{ShengliQM}  Another scaling relationship seen in heavy ion collisions is a uniform curve in the $v_{n}(p_{T})$ divided by the product of the initial state ellipticity, $\epsilon_{n}$, and the cube root of the number of participating nucleons, $N_{part}^{1/3}$, where both the $\epsilon_{n}$ and $N_{part}$ are determined from a Glauber calculation.~\cite{v2CuCu}  When the $0$-$5\%$ $d$+Au and $0$-$5\%$ $^{3}$He+Au $v_{2}$ results are scaled, giving $v_{2}/(\epsilon_{2} \times N_{part}^{1/3})$, and compared with the Au+Au measurements at a variety of centralities, a uniform curve is seen with the small systems results consistent with but slightly lower than the heavy ion values.  However, the $0$-$5$\% $^{3}$He+Au $v_{3}/(\epsilon_{3} \times N_{part}^{1/3})$ values are significantly below the uniform curve seen for the Au+Au centralities.  This scaling does not work as well in small collisions systems.  The failure of this scaling relationship in $v_{3}$ means that the higher-order anisotropies damp quickly in these small systems.~\cite{ShengliQM}

\begin{figure}
\begin{center}
\includegraphics[width=.6\textwidth]{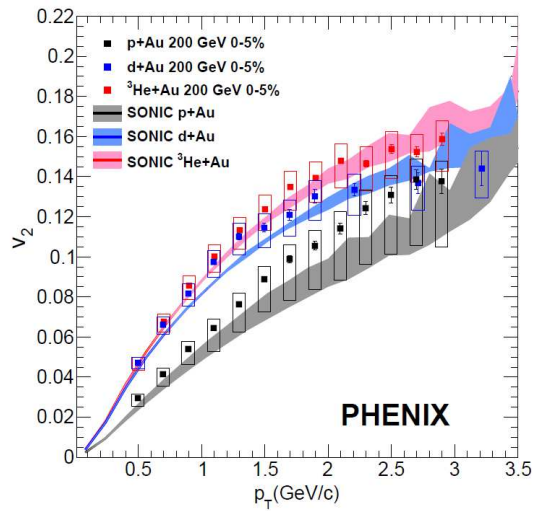}
\caption{The $v_{2}$ measured in $\sqrt{s_{NN}}=200$~GeV $p$+Au,~\cite{pAudeta} $d$+$Au$~\cite{dAudeta} and $^{3}$He+Au~\cite{He3Au} collisions at RHIC.  A comparison with calculations from the SONIC model is also shown.~\cite{SONIC}}
\label{Fig:comparev2}
\end{center}
\end{figure}

Long-range correlation measurements are also performed in the $^{3}$He+Au and $p$+Au systems.  Mid-rapidity hadrons are paired with photomultiplier hits in the BBC resulting in $|\Delta \eta|$ values between $2.75$ and $4.05$.  Figure~\ref{Fig:comparev2} shows a comparison of the $v_{2}$ values in $\sqrt{s_{NN}}=200$~GeV $p$+Au~\cite{pAudeta}, $d$+Au~\cite{dAudeta} and $^{3}$He+Au~\cite{He3Au} in the $0$-$5\%$ centrality bin.  A clear ordering of the $v_{2}$ values is seen with $v_{2}^{d+Au}$ and $v_{2}^{3He+Au}$ resulting in similar values and $v_{2}^{d+Au} > v_{2}^{p+Au}$.~\cite{SONIC}  This follows initial state eccentricity ordering where the $\epsilon_{2}^{d+Au}$ and $\epsilon_{2}^{3He+Au}$ are comparable and $\epsilon_{2}^{d+Au} > \epsilon_{2}^{p+Au}$.  These $v_{2}$ results are well reproduced by the SONIC model,~\cite{SONIC} while curves from an IP-Glasma initial state model with a hydrodynamic evolution~\cite{IPGlasma} fail to reproduce the data.

\section{Conclusion}

Initial state geometric effects are clearly seen in small system collectivity measurements at RHIC.  Evidence of this includes the $\epsilon_{2}$-ordering of the $v_{2}$ in $p$+Au, $d$+Au and $^{3}$He+Au collisions and the positive $v_{3}$ values measured in $^{3}$He+Au collisions.  At low $p_{T}$ the $v_{2}$ in $d$+Au and $^{3}$He+Au collisions exhibits mass-ordering and the $v_{2}$ in $^{3}$He+Au follows $n_{q}$-scaling at all $p_{T}$'s.  Low $p_{T}$ mass ordering and a dependence on the initial state collision geometry are both hallmarks of hydrodynamic behavior.  Additionally, the data is well reproduced by hydrodynamic models, in particular the SONIC model, which includes a hadronic cascade.  One interesting deviation seen in the $^{3}$He+Au $v_{3}$ is the lower values of $v_{3}/(\epsilon_{3} \times N_{part}^{1/3})$ suggesting a faster damping of $v_{3}$ in small systems.  A promising future measurement that is in progress is the $\pi^{0}$-MPC cluster correlation result which extends the measurement of these anistropies out to a higher $p_{T}$, up to $8$~GeV/$c$.


\begin{thebibliography}{99}
\bibitem{STAR} B. I. Abelev \emph{et al.}, \emph{Long range rapidity correlations and jet production in high energy nuclear collisions}, \emph{Phys. Rev. C} {\bf 80} (2009) 064912 [{\tt arXiv:0909.0191}].

\bibitem{PHOBOS} B. Alver \emph{et al.}, \emph{High transverse momentum triggered correlations over a large pseudorapidity acceptance in Au+Au collisions at $\sqrt{s_{NN}}=200$~GeV}, \emph{Phys. Rev. Lett.} {\bf 104} (2010) 062301 [{\tt arXiv:0903.2811}].

\bibitem{Luzum} M. Luzum, \emph{Collective flow and long-range correlations in relativistic heavy ion collisions}, \emph{Phys. Lett. B} {\bf 696} (2011) 499 [{\tt arXiv:1011.5773}].

\bibitem{GaleHydro} C. Gale, S. Jeon, B. Schenke, P. Tribedy, and R. Venugopalan, \emph{Event-by-event anisotropic flow in heavy-ion collisions from combined Yang-Mills and viscous fluid dynamics} \emph{Phys. Rev. Lett.} {\bf 110} (2013) 012302 [{\tt arXiv:1209.6330}].

\bibitem{GLASMA} K. Dusling and R. Venugopalan, \emph{Azimuthal collimation of long range rapidity correlations by strong color fields in high multiplicity hadron-hadron collisions}, \emph{Phys. Rev. Lett.} {\bf 108} (2012) 262001.

\bibitem{CMS} S. Chatrchyan \emph{et al.}, \emph{Observation of long-range near-side angular correlations in proton-lead collisions at the LHC}, \emph{Phys. Lett. B} {\bf 718} (2013) 795 [{\tt arXiv:1210.5482}].

\bibitem{ATLAS} G. Aad \emph{et al.}, \emph{Observation of associated near-side and away-side long-range correlations in $\sqrt{s_{NN}}=5.02$~TeV proton-lead collisions with the ATLAS detector}, \emph{Phys. Rev. Lett.} {\bf 110} (2013) 182302 [{\tt arXiv:1212.5188}].

\bibitem{ALICE} B. Abelev \emph{et al.}, \emph{Long-range angular correlations on the near and away side in $p$-Pb collisions at $\sqrt{s_{NN}}=5.02$~TeV}, \emph{Phys. Lett. B.} {\bf 719} (2013) 29 [{\tt arXiv:1212.2001}].

\bibitem{ppg149} A. Adare \emph{et al.}, \emph{Quadrupole anisotropy in dihadron azimuthal correlation in central $d$+Au collisions at $\sqrt{s_{NN}}=200$~GeV}, \emph{Phys. Rev. Lett.} {\bf 111} (2013) 212301 [{\tt arXiv:1303.1794}].

\bibitem{dAudeta} A. Adare \emph{et al.}, \emph{Measurement of long-range angular correlation and quadrupole anisotropy of pions and (anti)protons in central $d$+Au collisions at $\sqrt{s_{NN}}=200$~GeV}, \emph{Phys. Rev. Lett.} {\bf 114} (2015) 192301 [{\tt arXiv:1404.7461}].

\bibitem{TakaoQM} T. Sakaguchi for the PHENIX Collaboration, \emph{PHENIX results on centrality dependence of yields and correlations in $d$+Au collisions at $\sqrt{s_{NN}}=200$~GeV}, \emph{Nucl. Phys. A} {\bf 956} (2016) 300 [{\tt arXiv:1601.03450}].

\bibitem{LuzumRomatschke} M. Luzum and P. Romatschke, \emph{Conformal relativistic viscous hydrodynamics: Applications to RHIC results at $\sqrt{s_{NN}}=200$~GeV}, \emph{Phys. Rev. C}, {\bf 78} (2008) 034915.

\bibitem{SONIC} J. Nagle \emph{et al.}, \emph{Exploiting intrinsic triangluar geometry in relativistic $^{3}$He+Au collisions to disentangle medium properties}, \emph{Phys. Rev. Lett} {\bf 113} (2014) 112301 [{\tt arXiv:1312.4565}].

\bibitem{He3Au} A. Adare \emph{et al.}, \emph{Measurements of elliptic and triangular flow in high-multiplicity $^{3}$He+Au collisions at $\sqrt{s_{NN}}=200$~GeV}, \emph{Phys. Rev. Lett.} {\bf 115} (2015) 142301 [{\tt arXiv:1507.06273}].

\bibitem{Glauber} P. Bozek and W. Bronkowski, \emph{Hydrodynamic modeling of $^{3}$He+Au collisions at $\sqrt{s_{NN}}=200$~GeV}, \emph{Phys. Lett. B} {\bf 747} (2015) 135.

\bibitem{AMPT} Z.-W. Lin, C. M. Ko, B.-A. Li, B. Zhang, and S. Pal, \emph{A Multi-phase transport model for relativistic heavy ion collisions}, \emph{Phys. Rev. C}, {\bf 72} (2005) 064901.

\bibitem{superSONIC} P. Romatschke, \emph{Light-heavy collisions: A window into pre-equilibrium QCD dynamics?} \emph{Eur. Phys. J. C} {\bf 75} (2015) 305[{\tt arXiv:1502.04745}].

\bibitem{IPGlasma} B. Schenke and R. Venugopalan, \emph{Collective effects in light-heavy ion collisions}, \emph{Nuclear Phys. A} {\bf 931} (2014) 1039 [{\tt arXiv:1407.7557}].

\bibitem{ShengliQM} S. Huang for the PHENIX Collaboration, \emph{Measurements of elliptic and triangular flow in high-multiplicity $^{3}$He+Au collisions at $\sqrt{s_{NN}}=200$~GeV}, \emph{Nucl. Phys. A} {\bf 956} (2016) 761. 

\bibitem{v2CuCu} A. Adare \emph{et al.}, \emph{Systematic study of azimuthal anisotropy in Cu+Cu and Au+Au collisions at $\sqrt{s_{NN}}=62.4$ and $200$~GeV}, \emph{Phys. Rev. C} {\bf 92} (2015) 034913 [{\tt arXiv:1412.1043}].

\bibitem{pAudeta} C. Aidala \emph{et al.}, \emph{Measurement of long-range angular correlations and azimuthal anistropies in high multiplicity $p$+Au collisions at $\sqrt{s_{NN}}=200$~GeV} [{\tt arXiv:1609.02894}].


\end{thebibliography}
\end{document}